\documentclass[aps,prl,nofootinbib,amsmath,amssymb,twocolumn,10pt]{revtex4}%superscriptaddress,showpacs,

\pdfoutput=1

\usepackage{graphicx}
\usepackage{dcolumn}
\usepackage{bm}
\usepackage{amssymb}
\usepackage{latexsym}
\usepackage{booktabs}
\usepackage{amsmath}
\usepackage{multirow}
\usepackage[colorlinks=true, linkcolor=red, citecolor=blue]{hyperref}

\newcommand{\be}{\begin{equation}}
\newcommand{\ee}{\end{equation}}
\newcommand{\bq}{\begin{eqnarray}}
\newcommand{\eq}{\end{eqnarray}}

\begin{document}

\title{Probing the interaction between dark energy and dark matter with the parametrized post-Friedmann approach}

\author{Xin Zhang}%\footnote{Corresponding author}}
\email{zhangxin@mail.neu.edu.cn}
\affiliation{Department of Physics, College of Sciences,
Northeastern University, Shenyang 110004, China}
\affiliation{Center for High Energy Physics, Peking University, Beijing 100080, China}

\begin{abstract}
%This paper is a Research Highlight invited by SCIENCE CHINA Physics, Mechanics \& Astronomy. 

There possibly exists some direct, non-gravitational coupling between dark energy and dark matter. This possibility should be seriously tested by using observations, which requires us to understand such a scenario from the aspects of both expansion history and growth of structure. It is found that once calculating the perturbations in the interacting dark energy (IDE) scenario, for most cases the curvature perturbation on superhorizon scales is divergent, which is a catastrophe for the IDE cosmology. We found a solution to this issue, which is to establish an effective theory to treat the dark energy perturbations totally based on the basic facts of dark energy. This scheme generalizes the parametrized post-Friedmann framework of uncoupled dark energy and can be used to cure the instability of the IDE cosmology. The whole parameter space of IDE models can henceforward be explored by observational data. The IDE scenario can thus be tested or falsified with current and future observational data by using the PPF approach. We expect that the future highly accurate observational data would offer the certain answer to the question whether there is a direct coupling between dark energy and dark matter.

\end{abstract}

 \maketitle

In the current universe, the dominant energy components are dark energy and dark matter. There is a longstanding conjecture that there might be some direct coupling between dark energy and dark matter (for a recent review, see Ref.~\cite{Wang:2016lxa}). The advantages for considering such a possibility include, for example, those that it can alleviate the ``cosmic coincidence'' puzzle, can avoid the ``big rip'' in a phantom scenario, and so on. We call the scenario in which dark energy directly interacts with dark matter the ``interacting dark energy'' (IDE) scenario. Actually, besides the above reasons of theoretical aspect, one should be more concerned with the observational issue: How can we detect this interaction (this is essentially a kind of ``fifth force'') or falsify this scenario by using the observations? This requires us to be able to calculate how it affects the cosmological evolution, including both aspects of expansion history and growth of structure. 

However, once calculating the cosmological perturbation evolution of the IDE scenario, it was found that for most cases the curvature perturbation on superhorizon scales at early times is divergent, which is a catastrophe for the IDE cosmology \cite{Valiviita:2008iv}. This indicates that we actually do not know how to consider the perturbations of dark energy. If dark energy is not the cosmological constant $\Lambda$, then it cannot be treated as a pure background, which means that it also has perturbations as the response of the metric fluctuations. But, since we know little about the nature of dark energy, of course we do not understand how dark energy is perturbed. Undoubtedly, dark energy perturbation is an important issue for understanding the nature of dark energy.

The perturbation instability frustrated the advance of the IDE cosmology. This problem has to be solved if one wants to test the IDE scenario (or falsify it) by using the current and future observational data. My research group in the Northeastern University of China found a solution to this issue \cite{Li:2014eha}, which is to establish an effective theory to treat the dark energy perturbations totally based on the basic facts of dark energy. This scheme generalizes the parametrized post-Friedmann (PPF) framework of uncoupled dark energy \cite{Fang:2008sn} and can be used to cure the instability of the IDE cosmology \cite{Li:2014eha,Li:2014cee,Li:2015vla,Yin:2015pqa,Guo:2017hea}. 

In the traditional linear perturbation theory, we can write down the linear Einstein equations and for each energy component, we have the continuity and Navier-Stokes equations. Usually, for each component, we have three variables, $\delta\rho$, $v$, and $\delta p$, but we have only two equations to solve them. We thus need to find an additional equation to relate the pressure and density perturbations to complete the perturbation equations system. That is an equation for sound speed. However, for dark energy, there is a problem. Namely, the adiabatic sound speed $c_a$ is not physical because $c_a^2$ is less than zero. So, in this case, one has to impose by hand a physical sound speed for dark energy fluid, $c_s$, defined to be $c_s^2=\delta p/\delta \rho$ in the rest frame of dark energy ($v|_{\rm rf}=B|_{\rm rf}=0$). And, so we view dark energy as a nonadiabatic fluid. Then, in a general gauge, $\delta p$ has two parts---the adiabatic part and the nonadiabatic one. The interaction term will appear in the nonadiabatic part, and in some cases the nonadiabatic modes will lead to the instability.

\begin{figure*}[!htbp]
  \includegraphics[width=6cm]{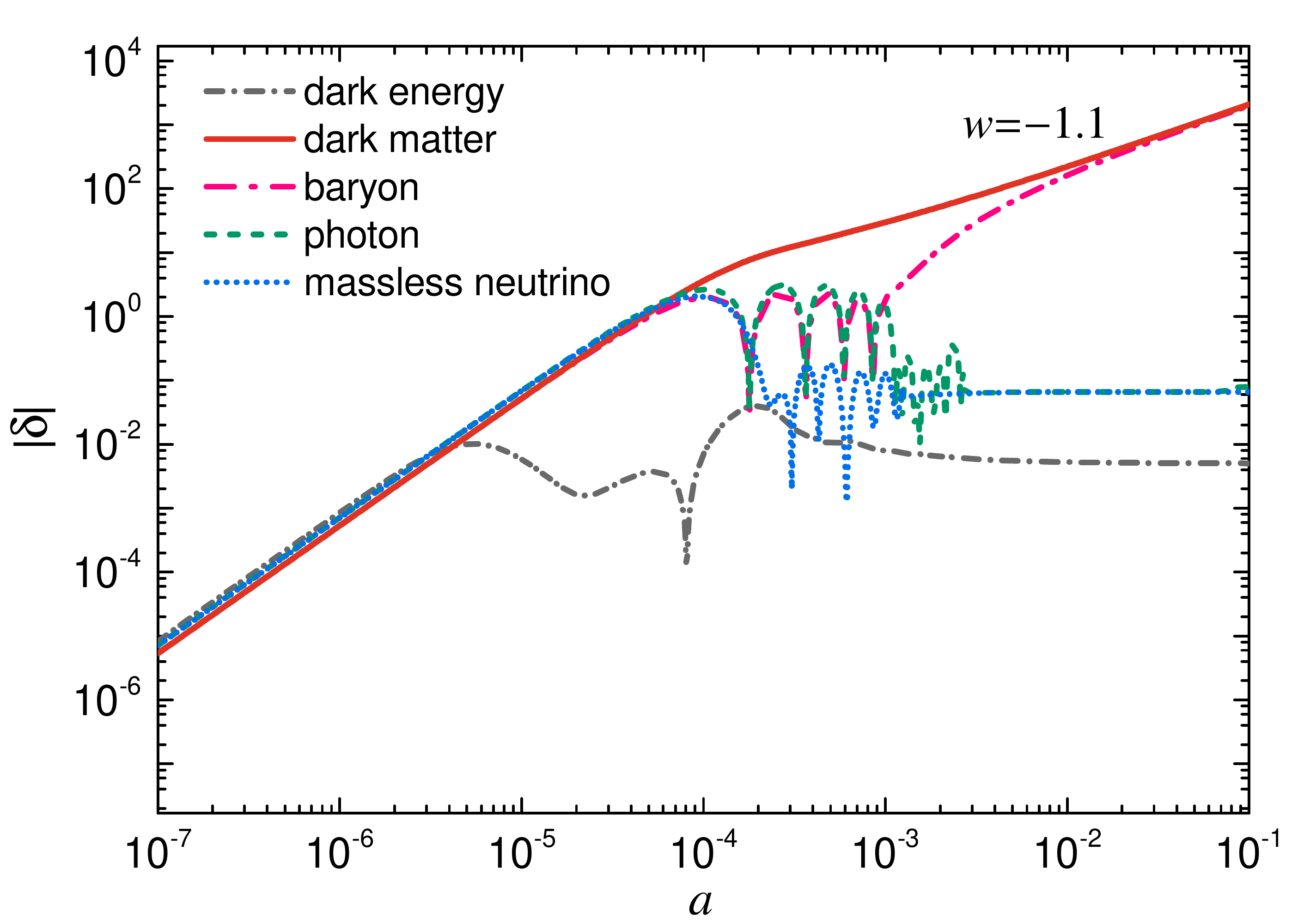}
  \includegraphics[width=6cm]{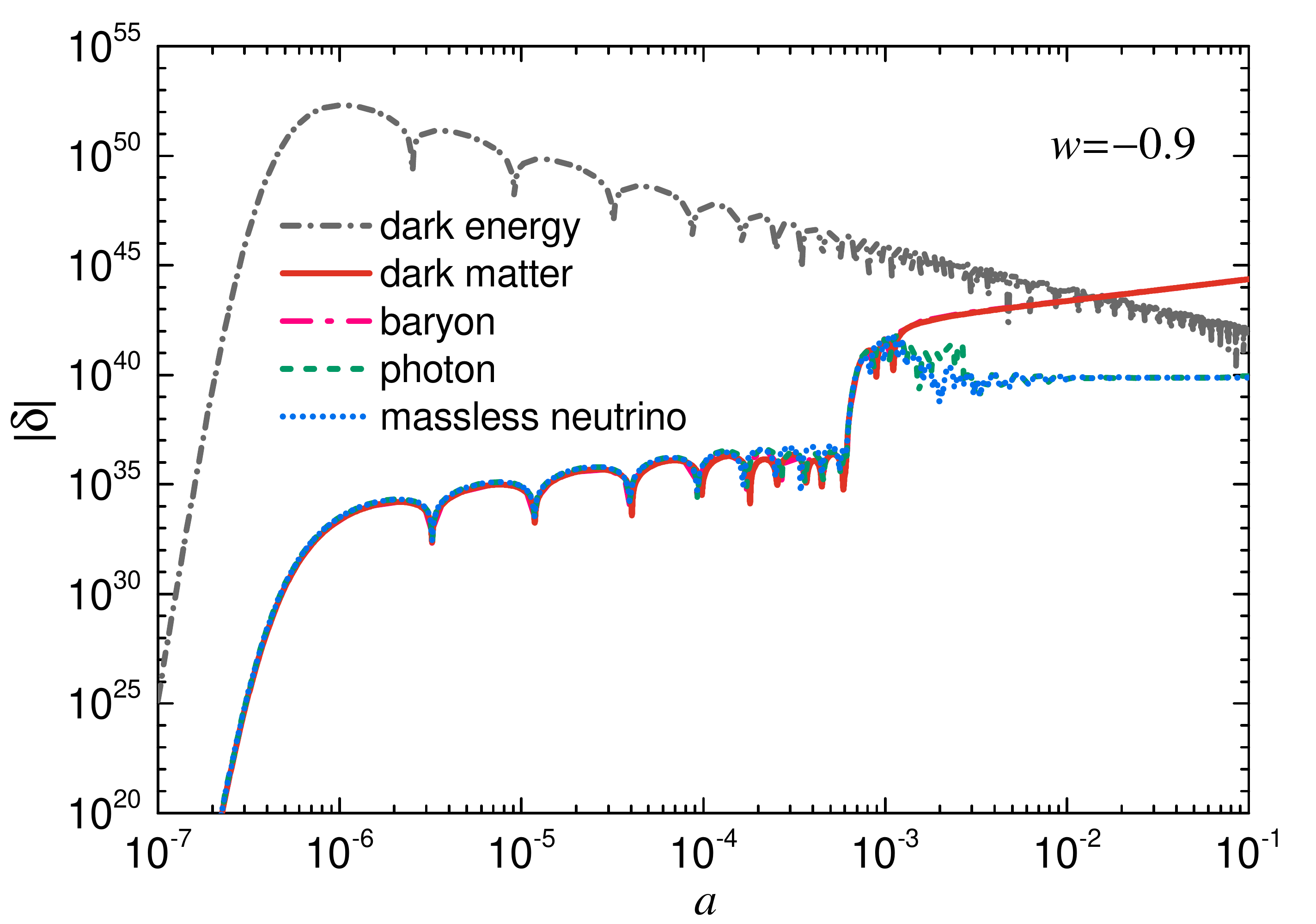}
  \includegraphics[width=6cm]{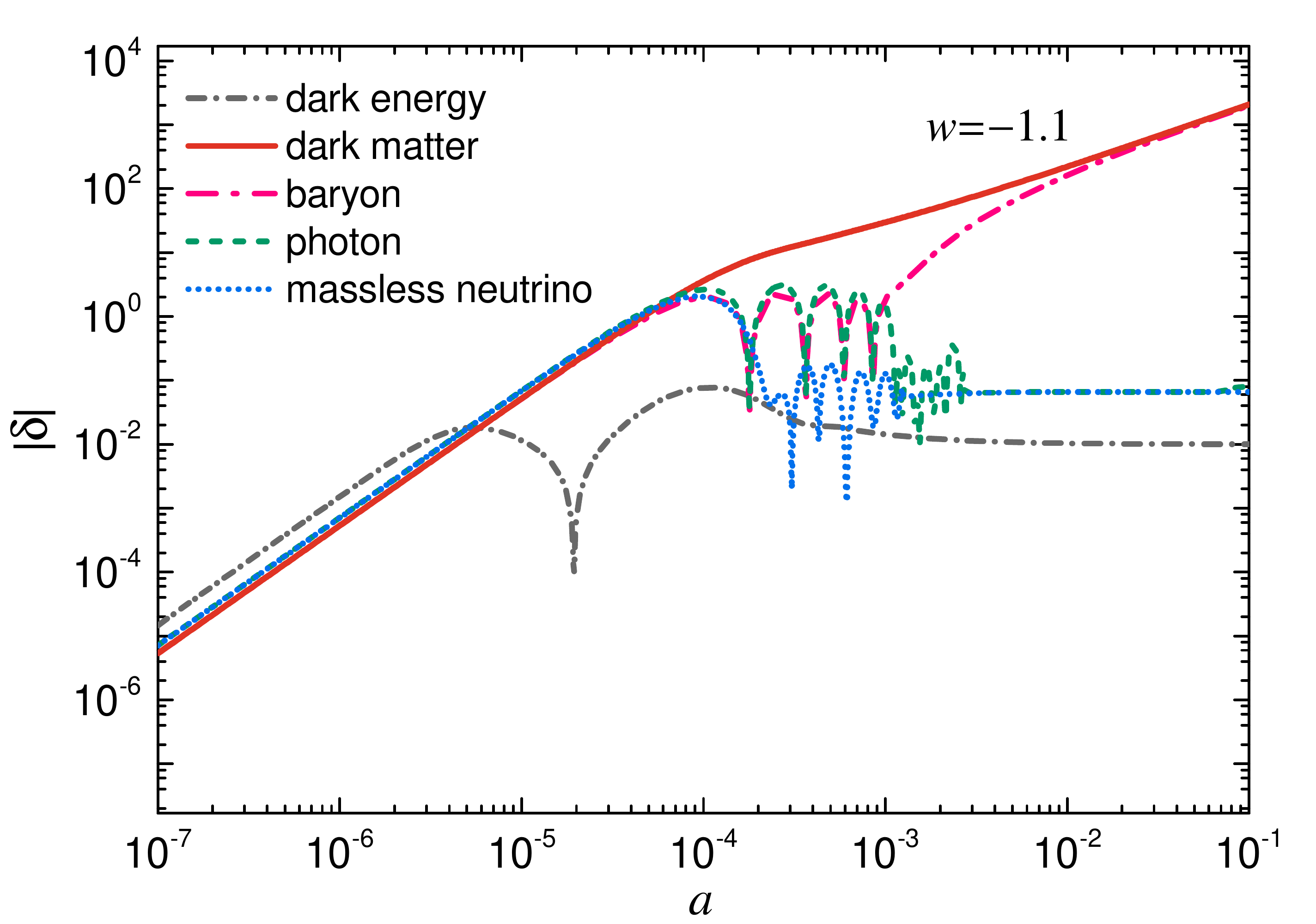}
  \includegraphics[width=6cm]{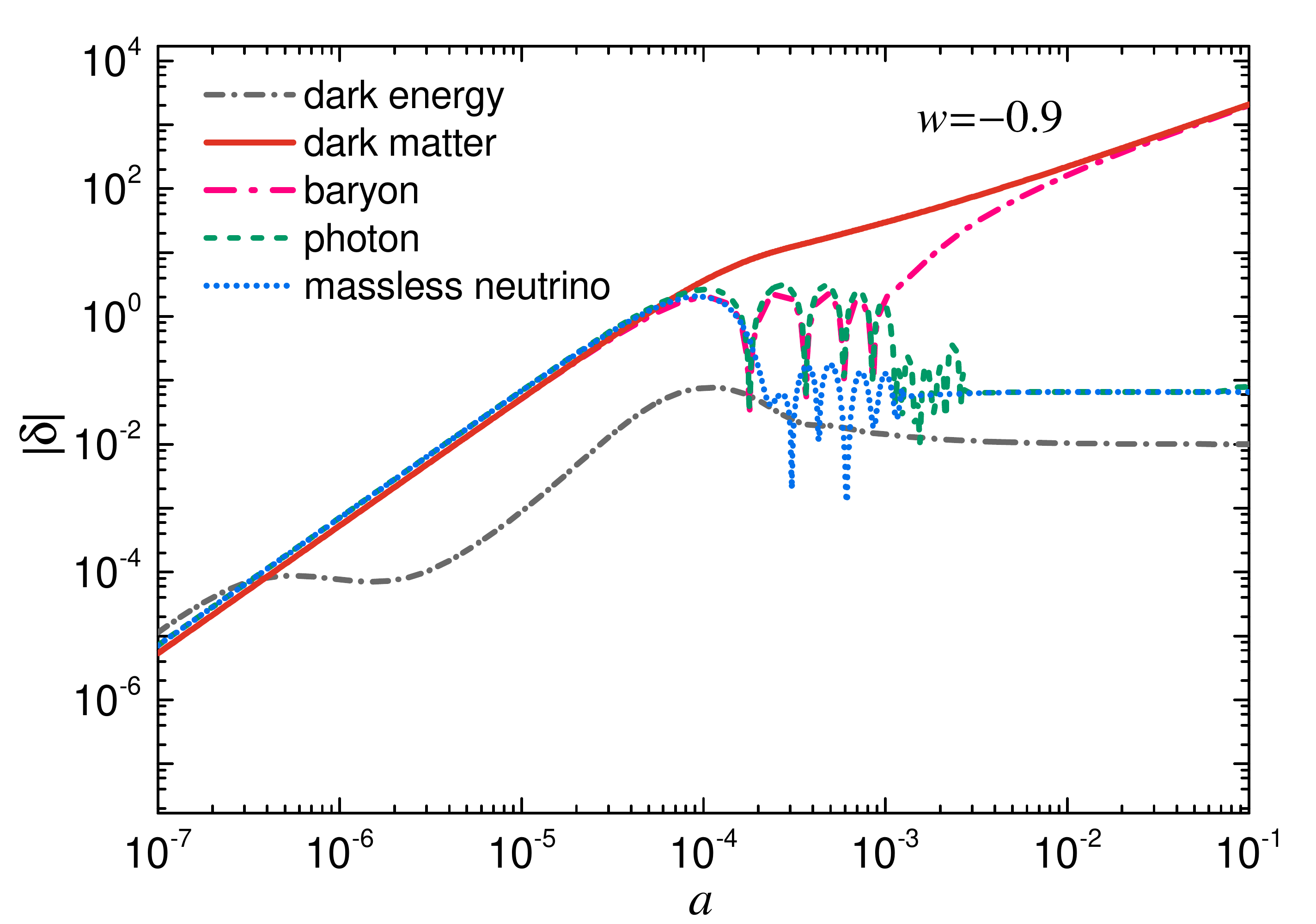}
  \caption{The density perturbation evolutions at $k=0.1~\rm{Mpc}^{-1}$ in the IDE model with $Q^\mu=3\beta H\rho_c u_c^\mu$ in the synchronous gauge. The upper panels are obtained by using the previous method, and the lower panels are obtained within the PPF framework. 
}\label{fig1}
\end{figure*}

\begin{figure}[!htbp]
  \includegraphics[width=7.5cm]{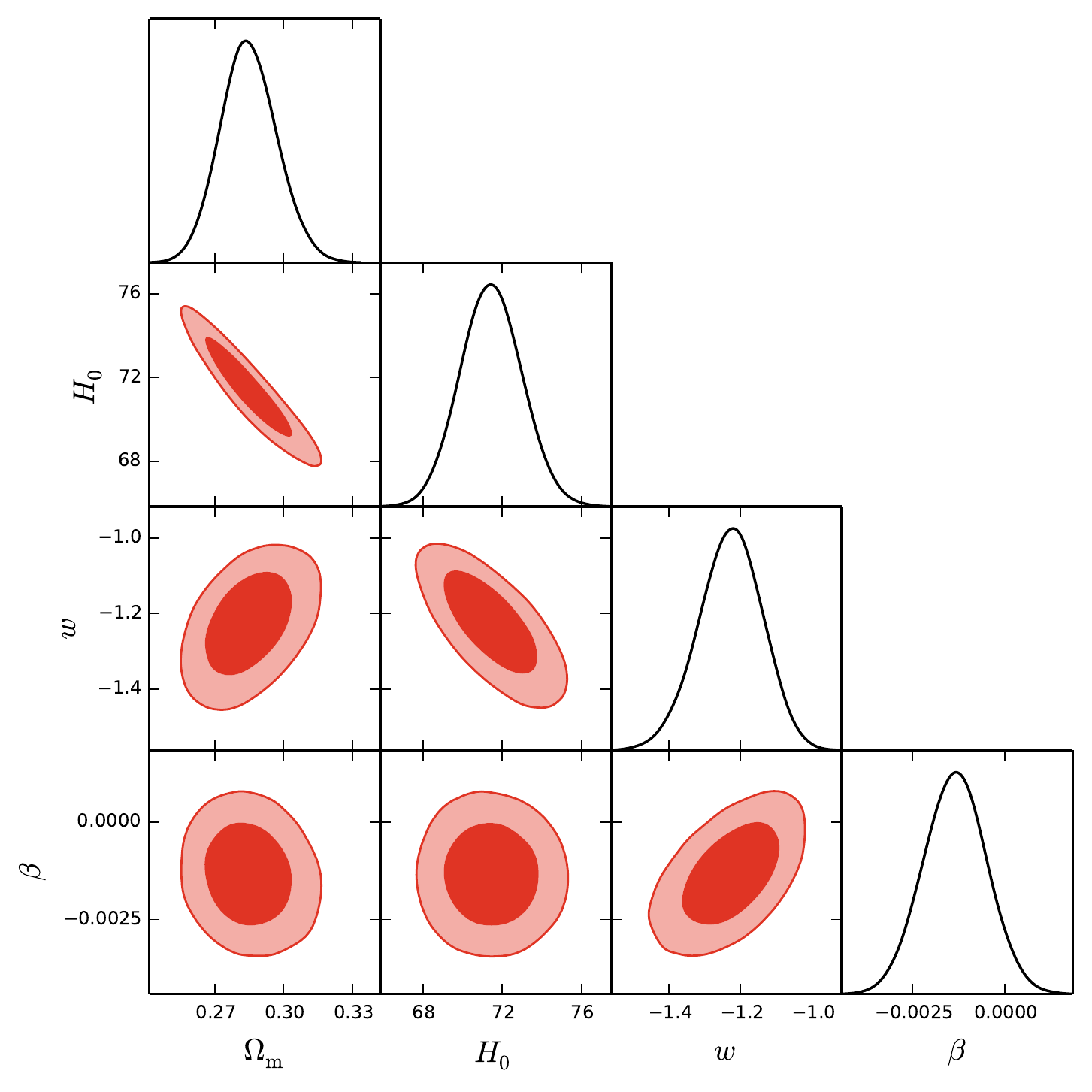}
  \caption{The one- and two-dimensional posterior distributions for the parameters 
  in the IDE model with $Q^\mu=3\beta H\rho_c u_c^\mu$.}\label{fig2}
\end{figure}

In 2008, Valiviita, Majerotto, and Maartens \cite{Valiviita:2008iv} found that once calculating the perturbations in IDE models, for most cases the curvature perturbation is divergent at early times. For the example of $Q\propto \rho_c$, they found that, when $w> -1$, the instability will appear, no matter how small the coupling is. The reason is that the pressure perturbation defined by sound speed involves nonadiabatic mode. The nonadiabatic mode in uncoupled dark energy models will decay away, but in IDE models sometimes will rapidly grow, leading to the blowup of curvature perturbation. This reveals our ignorance about the nature of dark energy. We actually do not know how to treat the perturbation of dark energy. In 2014, the author of the present paper and collaborators (Yun-He Li and Jing-Fei Zhang) \cite{Li:2014eha} found a solution to this issue: that is to establish an effective theory based on the basic facts of dark energy, which is the PPF framework. This generalizes the PPF of uncoupled dark energy of Fang, Hu, and Lewis \cite{Fang:2008sn}.

The PPF description does not consider the pressure perturbation of dark energy. It is totally based on the basic facts of dark energy. On large scales, far beyond the horizon, the relationship between the velocities of dark energy and other matters can be empirically parameterized. On small scales, deep inside the horizon, dark energy is smooth enough so that it can be viewed as a pure background; so we have the Poisson equation for this limit. In order to make these two limits compatible, we introduce a dynamical function $\Gamma$ by which we can find an equation to describe the cases on all scales. What we should do is to find the equation of motion on all scales for $\Gamma$ and finally this is successfully done. In the equation of motion of $\Gamma$, we introduce a parameter $c_\Gamma$ that gives a transition scale in terms of the Hubble scale under which dark energy is smooth enough. And, in this equation there is no perturbation variable of dark energy. Once the evolution of $\Gamma$ is derived, we can directly obtain the density perturbation and velocity perturbation of dark energy. So, this method avoids using the pressure perturbation defined by sound speed. We have shown that this new framework can give stable cosmological perturbations in the IDE scenario.

For example, for the IDE model with $w={\rm constant}$ and $Q^\mu=3\beta H\rho_c u^\mu_c$, the instability will occur when $w>-1$, within the framework of previous method (calculating the pressure perturbation of dark energy by imposing a physical sound speed by hand) \cite{Valiviita:2008iv}. We have shown that, when using the PPF scheme, the instability will be successfully cured \cite{Li:2014eha}. Figure \ref{fig1} plots the density perturbation evolutions at $k=0.1~\rm{Mpc}^{-1}$ for this model in the synchronous gauge. The upper panels are obtained by using the previous method, and the lower ones are obtained within the PPF framework. Typical examples are taken as $w=-1.1$ (left panels) and $w=-0.9$ (right panels), to show the cases of $w<-1$ and $w>-1$. 
We take the coupling strength $\beta=-10^{-17}$ and fix all the other parameters at their best-fit values from Planck. 
%Taking such a small value for $\beta$ is to avoid the possible breakdown of the numerical computation when the instability occurs in the IDE model using the old method.
We can clearly see that, once we use the PPF method (lower panels), the instability can be avoided. The cosmological perturbations are stable for both cases of $w<-1$ and $w>-1$. This example demonstrates that the PPF scheme for IDE scenario works very well. Now the whole parameter space of the IDE model can be explored by using the observational data. The constraint results for this model are shown in Fig. \ref{fig2}, where the current CMB+BAO+SN+$H_0$ observational data are used. Note also that here Figs. \ref{fig1} and \ref{fig2} are taken from Ref.~\cite{Li:2014eha}.

After a careful test, we conclude that the new PPF framework is applicable to all the IDE models. Also, this new PPF scheme is downward compatible with the previous one for uncoupled dark energy. Using this PPF scheme for IDE, one can calculate the perturbation evolution and structure growth in all IDE models, and further use the measurements of growth of structure, such as weak lensing, galaxy cluster number counts, and redshift space distortions, combined with the measurements of expansion history, to constrain the IDE models. In particular, the whole parameter space of IDE models can, henceforward, be explored by using the observations.

We end this paper by a brief summary. There possibly exists some direct, non-gravitational coupling between dark energy and dark matter. This possibility should be seriously tested by using observations, which requires us to understand such a scenario from the aspects of both expansion history and growth of structure. It is found that once calculating the perturbations in the interacting dark energy scenario, for most cases the curvature perturbation on superhorizon scales is divergent, which is a catastrophe for the IDE cosmology. The author of the present paper and collaborators found a solution to this issue, which is to establish an effective theory to treat the dark energy perturbations totally based on the basic facts of dark energy. This scheme generalizes the parametrized post-Friedmann framework of uncoupled dark energy and can be used to cure the instability of the IDE cosmology. The whole parameter space of IDE models can henceforward be explored. The IDE scenario can thus be tested (or falsified) with current and future observational data by using the PPF approach. We expect that the future highly accurate observational data would offer the certain answer to the question whether there is a direct coupling between dark energy and dark matter.

\begin{acknowledgments}
This work was supported by the National Natural Science Foundation of China (Grants No.~11522540 and No.~11690021), the Top-Notch Young Talents Program of China, and the Provincial Department of Education of Liaoning (Grant No.~L2012087).

\end{acknowledgments}

\end{document}